\documentclass[onecolumn, 12pt]{article}

\usepackage{cite}

\usepackage[cmex10]{amsmath}

\usepackage{epsfig}

\usepackage{amssymb}

\begin{document}

\title{On the Mobile-to-Mobile Linear Time-Variant\\Shallow-Water Acoustic Channel Response\footnote{This work is under review for Journal publication. Copyright may be transferred without notice, after which this version may no longer be accessible.}~\footnote{This work is partially supported by the Spanish Government and FEDER under project TEC2014-57901-R and the Junta de Andaluc\'{i}a under projects P11-TIC-7109 and P11-TIC-8238.}}
%
%
%

\author{Adri\'{a}n Sauco-Gallardo,
        Unai Fern\'{a}ndez-Plazaola,
        and Luis D\'{i}ez\footnote{The authors are with the Department of Communication Engineering, University of M\'{a}laga International Campus of Excellence Andalucía TECH, 29071 Málaga, Spain (e-mail: asaucog@ic.uma.es).}}

\maketitle

\begin{abstract}
We expose some concepts concerning the channel impulse response (CIR) of linear time-varying (LTV) channels to give a proper characterization of the mobile-to-mobile underwater channel. We find different connections between the linear time-invariant (LTI) CIR of the static channel and two definitions of LTV CIRs of the dynamic mobile-to-mobile channel. These connections are useful to design a dynamic channel simulator from the static channel models available in the literature. Such feature is particularly interesting for overspread channels, which are hard to characterize by a measuring campaign. Specifically, the shallow water acoustic (SWA) channel is potentially overspread due the signal low velocity of propagation which prompt long delay spread responses and great Doppler effect. Furthermore, from these connections between the LTI static CIRs and the LTV dynamic CIRS, we find that the SWA mobile-to-mobile CIR does not only depend on the relative velocity between transceivers, but also on the absolute velocity of each of them referred to the velocity of propagation. Nevertheless, publications about this topic do not consider it and formulate their equations in terms of the relative velocity between transceivers. We illustrate our find using two couples of examples where, even though the relative velocity between the mobiles is the same, their CIRs are not.
\end{abstract}


\section{Introduction}

Underwater acoustic (UWA) channels is a developing research subject since the end of the eighties when it aroused great interest in applications such as collection of data for oceanic research, telemetry for pollution monitoring, offshore oil industry control, and remote control of underwater unmanned vehicles \cite{uac_concepts}. For this reason several works analyzing the UWA channel have emerged ever since \cite{uac_networks, variabilidad_shallowwater, stojanovic_geometria, stojanovic_velocidadrelativa2, zajic_mimo, overspread,swa_2order}.

Mobile-to-mobile (M2M) channels are a classic example in comunications of a linear time-varying (LTV) system which have been extensively studied \cite{andrea}. In radio channels, where electromagnetic waves are commonly employed, it is standard to separate the \emph{slow} variations from the \emph{fast} ones for the analysis, assuming the \emph{slow} variations contribution as a constant mean when separately analyzing the \emph{fast} ones. This assumption is valid when the channel is not overspread, i.e. the product of the delay spread and the maximum Doppler shift is much smaller than one.

The M2M shallow water acoustic (SWA) channel suffers from harsh multipath propagation, caused by strong reflections on the seabed and water surface, together with the low velocity of propagation of acoustic waves ($c\sim 1500$ m/s). The combination of these two characteristics causes extremely long delay spread. The low velocity of propagation also is the cause of extreme Doppler effect even for not really fast motions of the transceivers \cite{stojanovic_geometria,stojanovic_velocidadrelativa2, zajic_mimo, overspread, swa_2order}. Thus, we have to infer that this kind of channel is very likely overspread. These differences with common wireless channels encouraged us to study the M2M SWA channel with distinct rigor, since classic wireless channel assumptions for analysis no longer apply. Thus, our analysis focus on getting the M2M channel impulse response (CIR) without falling into a misuse of preconceived concepts.

In the literature it can be found several SWA static channel models, which have proved its validity for sundry real scenarios in stationary conditions. They have different levels of complexity, although they all are a geometry-based ray tracing model: from simpler deterministic static models as the one proposed in \cite{stojanovic_geometria, stojanovic_velocidadrelativa2} to the ones which add the random effect of water surface waves and underwater displacements or the scattered \emph{micropaths} around the predefined \emph{eigenpaths} because of imperfect reflections \cite{zajic_mimo,swa_2order}.  They all use the frequency- and length-dependent absorption loss to define each path as a low-pass response with a different amplitude and time delay. The responses of each path summed together form the channel response.

When we allow for the motion at will of the transmitter (Tx) and/or the receiver (Rx), we find a LTV channel, which is no longer stationary. We here propose two system structures inspired by \cite{ltv_structures} to give two different definitions of the LTV CIR in terms of the static or stationary responses given in the literature, which are treated as a spatial sampling along the geometry described by the mobiles. These two definitions enabled us to construct a simulator of this kind of LTV channels, becoming each of them a more straightaway option depending on how is the motion to consider.

Many of the publications already mentioned also discuss this topic; however, they do not give explicit details about how their LTV CIR are defined and give them in terms of the relative motion of the transceivers. Because of this, they may result misleading since when using mechanical waves, i.e. they propagate through a medium, the LTV response does not only depend on the relative Tx-Rx velocity, but on each of their velocities referred to the wave propagation speed in the medium. We must recall that the reciprocity approximation assumption requires that the mobiles velocities are very small compared to the speed of propagation \cite{ref_Dopplergeneral}. 

We show this with examples in a homogeneous medium where the motion of the transceivers takes place in a plane. Specifically an example where the receiver is moving away from a still transmitter at constant speed and the one where the transmitter is the one moving away from a still receiver at the same speed; and another example which compares the case where the transceivers are moving at the same velocity in the same direction so they keep the same distance within time and the static case where they are not moving and they simply are at the same distance than that in the latter dynamic case. In these two pair of examples the relative motion of the transceivers is the same, however we show the LTV response is not the same.

The remainder of this paper is organized as follows. Section II presents different ways to characterize a LTV system. In Section III we show how the introduced in the previous section applies to mobile-to-mobile SWA channels together with some remarkable examples. Then, in Section IV, we discuss some numerical results on the particular examples presented in previous section which enlighten the need for the work here reported. Finally, the conclusions of our work are discussed in Section V.

\section{Linear Time-Variant Channel Impulse Responses}\label{ltvcirs}

A LTV system is fully characterized by its Green's function $g(n,m)$ \cite{ltv_structures}, \cite[Section~3.5.1]{multirate_dsp}, which represents the system response at time $n$ to an impulse at time $m$. By employing the linear systems superposition principle we can calculate the LTV system output $y(n)$ to an input $x(n)$ as

\begin{equation}
y(n)=\sum_m g(n,m) x(m).
\end{equation} 

\begin{figure}[!t]
\centering
\includegraphics[width=0.85\columnwidth]{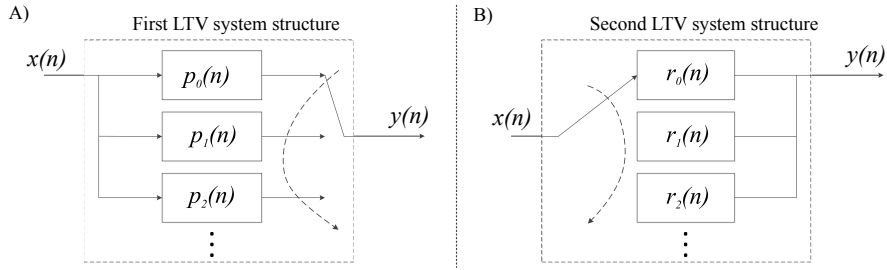}
\caption{A) First LTV system structure. The output switches every $n$ to the next LTI system with response $p_n(m)$. B) Second LTV system structure. The input switches every $n$ to the next LTI system with response $r_n(m)$.\label{LTVstructures12}}
\end{figure}

From Green's function we can define

\begin{equation}
p_n(m)=g(n,n-m),
\end{equation}
which is the system response at $n$ caused by an impulse input $m$ instants before, i.e. at $(n-m)$. Therefore, we can also express the LTV system output as

\begin{equation}
y(n)=\sum_m p_n(m) x(n-m),
\end{equation}
i.e. a convolution. Hence we obtain a first LTV system structure: the system output at each $n$ corresponds to the output of a different LTI system whose CIR is $p_n(m)$ and the input is always $x(n)$ as shown in Fig. \ref{LTVstructures12}.A. We name $p_n(m)$ the type I LTV CIR. This type I LTV CIR is the most commonly encountered when discussing LTV systems even though it is not properly a CIR. By what we mean that this is not the response to a given impulse but to one which must be placed at $(n-m)$ to get to know the response at $n$.

On the other hand, we can also define from Green's function
\begin{equation}
r_n(m)=g(n+m,n),
\end{equation}
which is the response $m$ instants later to a given impulse at $n$, therefore this is properly a CIR. The system output can also be expressed as

\begin{equation}
y(n)=\sum_m r_m(n-m) x(m),
\end{equation}
which is also a convolution. Thus we obtain the second LTV system structure: the system output corresponds to the superposition of the LTI systems outputs whose CIR is $r_n(m)$ and the input to each one has been one of the samples of the input $x(n)$ as shown in Fig. \ref{LTVstructures12}.B. We name $r_n(m)$ the type II LTV CIR.

These two structures are inspired on the periodically time-varying ones shown in \cite{ltv_structures}.

The connection between the two LTV CIR types can be straightaway derived from their relationship with Green's function:

\begin{align}
&p_n(m)=g(n,n-m)=r_{n-m}(m)\label{eq_pr}\\
&r_n(m)=g(n+m,n)=p_{n+m}(m)\label{eq_rp}.
\end{align}

\section{Application of the LTV CIRs for M2M SWA Channels}\label{applications}

In the SWA medium, if we consider a possible motion pattern of the Tx and/or the Rx, we obtain a LTV system which can be characterized using the structures explained in previous section, 

\begin{align}
p_n(m)=&h^\text{S}_{a(n-m),b(n)}(m)\\
r_n(m)=&h^\text{S}_{a(n),b(n+m)}(m),
\end{align}
where $a(n)$ and $b(n)$ denote the position of the Tx and the Rx at $n$, the superscript $\text{S}$ denotes static, and $h^{\text{S}}_{a,b}(m)$ is the CIR of the SWA static channel between points $a$ and $b$. For each $h^{\text{S}}_{a,b}(m)$, we can employ the expressions some accepted geometrical model of the SWA static channel available in the literature like \cite{stojanovic_geometria} or \cite{zajic_mimo}.  For a better understanding of these expressions we recall that type I LTV CIR $p_n(m)$ is the channel response at time $n$ to an impulse transmitted at time $n-m$, thus we must pay attention to where the Rx is at $n$ and where the Tx was at $n-m$. On the other hand, using $r_n(m)$ we must look at the Rx at $n+m$ and the Tx at $n$. These equations result from the propagation medium immobility, so that every spatial and velocity reference is taken from it, unlike in electromagnetic scenarios.

For the sake of simplicity, though without loss of generality, we developed our work from now on using the SWA channel model proposed and corroborated in \cite{stojanovic_geometria}. In this model the medium is homogeneous in all directions, the boundary conditions, i.e. seabed and surface, present neither temporal nor spatial variation and they are flat and parallel. The model allows for an easy computation of the channel frequency response (CFR) at any frequency in our band of operation, $H^\text{S}_{a,b}(f)$, and indirectly the CIR, $h^\text{S}_{a,b}(n)=\mathcal{F}^{-1}\{H^\text{S}_{a,b}(f)\}$. The superscript $\text{S}$ refers to static and $\mathcal{F}^{-1}\{\cdot\}$ is the inverse discrete Fourier transform operator. We will employ this model for our numerical section.

\subsection{Particular cases}\label{particularcases}

\begin{figure}[!tb]
\centering
\includegraphics[width=0.85\columnwidth]{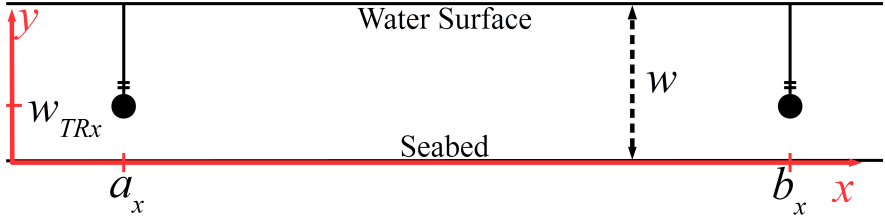}
\caption{Geometry of the studied scenarios.\label{escenario}}
\end{figure}

In this subsection we underline the difference between LTV CIR types by applying them to some illustrative scenarios. Fig. \ref{escenario} shows a basic simple scenario where we settle this cases. For the sake of simplicity, we will only consider motion of the transceivers contained in the same plane defined by the Cartesian axes $x$ and $y$, where we also consider that the scenario is invariant within $x$. Thus, this two-dimensional figure represents the whole geometry of our scenario. The total depth from the water surface to the seabed is $w$, while $a(n)$ and $b(n)$ are the points that represent the location of the Tx and the Rx at time $n$ respectively. Each point is given by a pair of Cartesian coordinates referred to the axes $x$ and $y$ shown in red color. For even more simplicity, we will also restrict the motion of the transceivers to straight-line and parallel to the seabed at the same depth, i.e. $a(n)=(a_x(n),w_{\text{Tx}})$, $b(n)=(b_x(n),w_{\text{Rx}})$, $w_{\text{Tx}}=w_{\text{Rx}}=w_{\text{TRX}}$. Therefore, {the static model response \cite{stojanovic_geometria}} only depends on the  distance $d$ between the transceivers, $h^\text{S}_{a,b}(m)=h^\text{S}_{||a-b||}(m)=h^\text{S}_{d}(m)$.

\subsubsection{Still Tx with moving Rx}\label{sss_Rx}

In the particular case the Tx is not moving at all, we can write $a_x(n)=a_0, \forall n$. Thus we can write the type I CIR of the LTV channel as

\begin{equation}\label{eq_pRx}
p^{\text{Rx}}_n(m)=h^\text{S}_{||a(n-m)-b(n)||}=h^\text{S}_{|a_0-b_x(n)|}(m).
\end{equation}
where the superscript $\text{Rx}$ states that is the Rx the one moving.

\subsubsection{Still Rx with moving Tx}\label{sss_Tx}

On the other hand, if it is the Rx the one standing still, we can write $b_x(n)=b_0, \forall n$. Hence, the type II CIR of this LTV channel is

\begin{equation}\label{eq_rTx}
r^{\text{Tx}}_n(m)=h^\text{S}_{||a(n)-b(n+m)||}=h^\text{S}_{|a_x(n)-b_0|}(m),
\end{equation}
where the superscript $\text{Tx}$ states that is the Tx the one moving.

It is inmediate to notice that we can have $a_0, b_x(n)$ from (\ref{eq_pRx}) and $b_0, a_x(n)$ from (\ref{eq_rTx}) such that the relative positions between transceivers within time is the same in both examples, ${|a_0-b_x(n)|=|a_x(n)-b_0|}$. As a simple example we give the one where we have a constant speed, $v$, departing from a distance $d_0$ at $n=0$ for both cases:

\begin{align}
d^{\text{Rx}}(n)=&| a_0 - b_x(n) |=|a_0 - (a_0+d_0+vn)| = d_0+vn,\\
d^{\text{Tx}}(n)=&| a_x(n) - b_0 |=| (a_0-vn)- (a_0+d_0) | = d_0+vn,\\
d^{\text{Rx}}(n)=&d^{\text{Tx}}(n)=d(n)=d_0+vn.
\end{align}

Nevertheless, the LTV channels of each example are different to each other, since $h^\text{S}_{d(n)}(m)$ corresponds to different type of LTV CIR depending on which transceiver is moving,

\begin{equation}\label{eq_basica}
p^{\text{Rx}}_n(m)=r^{\text{Tx}}_n(m)=h^\text{S}_{d(n)}(m),
\end{equation}
i.e. the type I CIR in the moving Rx case is the same to the type II CIR in moving Tx case and the temporal evolution along $n$ of both corresponds to each of the CIRs of a static SWA channel with separation between transceivers $d(n)$.

The reasons behind these results can be comprehended by thinking that when the Rx is moving, it determines the distance $d(n)$ at each $n$. Thus, $p_n(m)$, which is the response of the varying channel at $n$, corresponds to the response of the channel when the distance is $d(n)$, $h^\text{S}_{d(n)}(m)$. While when the Tx is the one moving and, hence, determining the distance $d(n)$ at each $n$, it is $r_n(m)$, which is the response to an impulse transmitted at $n$, the type which corresponds to $h^\text{S}_{d(n)}(m)$.

By using the relations between the type I and II responses stated in (\ref{eq_pr}), (\ref{eq_rp}), we also have

\begin{equation}\label{eq_r_Rx}
r^{\text{Rx}}_n(m)=p^{\text{Rx}}_{n+m}(m)=h^\text{S}_{d(n+m)}(m),
\end{equation}
\begin{equation}
p^{\text{Tx}}_n(m)=r^{\text{Tx}}_{n-m}(m)=h^\text{S}_{d(n-m)}(m).
\end{equation}

Hence we obtain that, despite the fact that the relative motions are the same, the LTV channels are not the same since their CIRs are not the same when expressing them by means of the same type of LTV CIR for both.

\subsubsection{Static case versus Tx and Rx moving keeping a constant distance}\label{sss_SvsD}

Now we compare a static case, $a_x(n)=a_0, b_x(n)=a_0+d_0$, with other dynamic case where both transceivers are moving with constant velocity although keeping the same distance as in the static case within time, $a_x(n)=a_0+vn, b_x(n)=a_0+d_0+vn$. Hence in both cases $d(n)=d_0, \forall n$.

The static case is simple and we write its LTI response,

\begin{equation}\label{eq_static}
h(m)=h^\text{S}_{d_0}(m).
\end{equation}

In the dynamic case, the channel is always the same despite the motion, nevertheless let us formulate it as a LTV system using, for instance, the type I CIR,

\begin{equation}
p_n(m)=h^\text{S}_{||a(n)-b(n)||}=h^\text{S}_{| (a_0+v(n-m))-(a_0+d_0+vn) |}(m)=h^\text{S}_{d_0+vm}(m),
\end{equation}
where we can observe that the LTV CIR loses its dependence on time $n$. Hence, this dynamic case is LTI as expected. If we choose the type II LTV CIR to formulate this case, it is easy to prove that the same expression is reached, since it is a LTI channel. Therefore, we write the channel response using the LTI CIR notation,

\begin{equation}\label{eq_dynam_LTI}
h^\text{D}(m)=h^\text{S}_{(d_0+vm)}(m),
\end{equation}
where the superscript $\text{D}$ stands for dynamic.

Although the mobile scenario turns out to be LTI, its response is different to the one of the static case. The different CIR of the mobile scenario is the result of the fact that the later a reflection arrives the further will have the Rx moved away, i.e. looking at (\ref{eq_dynam_LTI}) the later delay (the independent variable of the function, $m$) the larger is the distance of the channel (the subscript of the function, $d_0+vm)$.

\section{Numerical results} \label{numericalresults}

Here, we present numerical results for the particular cases detailed in \ref{particularcases}. We show some plots of the LTV and LTI CIRs with specific figures.

The channel model used is the deterministic one proposed in \cite{stojanovic_geometria} in the band of up to $128$ kHz. The depth of the water is $w=18$ m, the Tx and Rx are both $w_{\text{TRx}}=12$ m above the seabed, and the starting distance between them is $d_0=100$ m. As in \cite{stojanovic_geometria}, we will consider the speed of sound in the bottom $c_b=1300$ m/s and the density $\rho_b=1800$ g/m$^3$ to calculate the reflection coefficients. The constant velocity $v$ considered is $51.2$ m/s ($\approx 99.5$ knots), which is a fast one (although such can be found in torpedos) so the effects we want to show are noticeable at a glance.

\subsection{Still Tx versus still Rx}

We consider now the scenarios detailed in \ref{sss_Rx},\ref{sss_Tx}, i.e. the different LTV CIR we obtain when it is either the Rx or the Tx the one in motion at $v=51.2$ m/s.

\begin{figure}[!t]
\centering
\includegraphics[width=1\columnwidth]{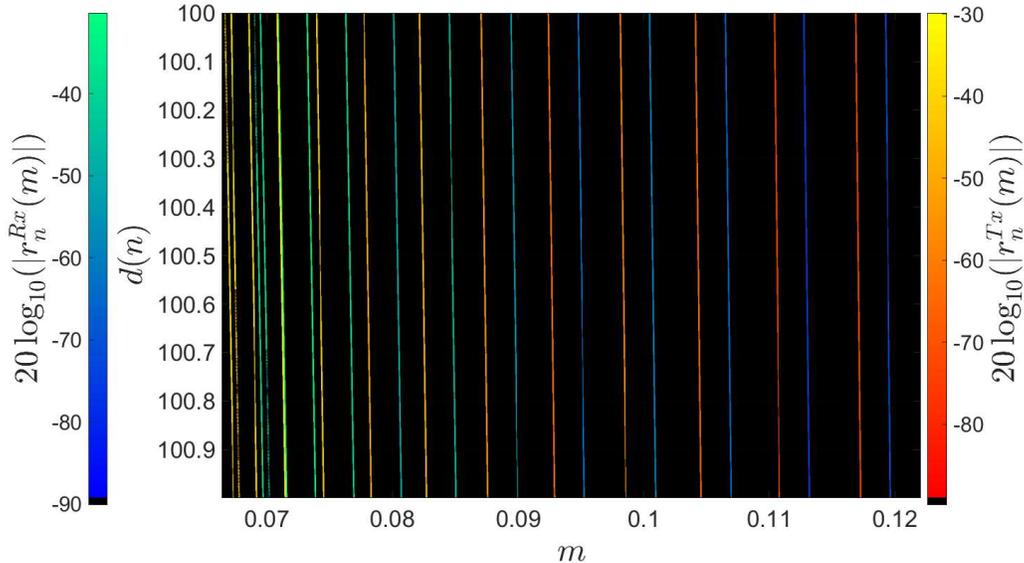}
\caption{Type II LTV responses, $r^{\text{Rx}}_n(m), r^{\text{Tx}}_n(m)$, for scenarios detailed in \ref{sss_Rx},\ref{sss_Tx}. In these examples $v=51.2$ m/s, $w=18$ m, $w_{\text{TRx}}=12$, $c_b=1300$ m/s, $\rho_b=1800$ g/m$^3$. The horizontal axis represents the evolution at $m$ (in seconds), the vertical axis represents the $d(n)$ (in meters). The magnitude of $r^{\text{Rx}}_n(m)$ is depicted by a cool color scale whereas the magnitude of $r^{\text{Tx}}_n(m)$ is depicted by a warm color scale.\label{fig_particular1_tipoII}}
\end{figure}

In Fig. \ref{fig_particular1_tipoII} the type II CIR, $r_n(m)$, is depicted for both cases within the first meter of the trajectory, i.e. from $100$ to $101$ m. We observe how the responses are similar but with a time shift when the Rx is the one moving away.

An explanation for this phenomenon can be found if we think on the definition of $r_n(m)$, the response at $n+m$ to the impulse sent at $n$: When the Tx is moving away, it sends an impulse at $n$, so the different reflected rays travel the geometry fixed by the distance $d(n)$ despite the Tx moving further and further away. On the other hand, when the Rx is the one moving away, the distance between the transceivers for the reflected ray at $n+m$ keeps growing as the Rx is traveling away. 

Let us now study the magnitude of the time shifts between the different cases. We first observe the cross-section at ${d(n=0)=d_0}$. It is immediate that the first arriving ray on the moving Tx case corresponds to the line of sight (LOS) component, its delay is $\tau^{\text{Tx}}_0|_{n=0}=d_0/c=66.66$ ms. Whereas on the moving Rx case, the velocity of the Rx moving away from the Tx acts as the effect of a reduction in the effective propagation speed, $\tau^{\text{Rx}}_0|_{n=0}=d_0/(c-v)= 69.02$ ms. Thus, we have that the moving Rx causes a time shift of approximately $2.36$ ms. The same time shift can also be observed between all the reflected rays of one case and their counterparts of the other. As the delays of the different rays of each case correspond to the rays traveling longer vertical distances (yet same horizontal distance) because of the zig-zag propagation, we can infer that the Rx horizontal movement causes a reduction of propagation speed that only affects to the horizontal propagation since the time shift is the same between all the rays of the two cases. This reduction of relative propagation speed also reduces the Doppler effect from one case to the other with a factor $(c+v)/(c-v)$, as we can infer from the general Doppler equation \cite{ref_Dopplergeneral}, 

\begin{equation}
f_D=\frac{(c-v_{\text{Rx}})}{(c+v_{\text{Tx}})}f_0.
\end{equation}

On the other hand, if we focus now on the effect caused as the time $n$ goes by and $d(n)$ grows, we observe how all the delays shift to the right as there is a longer path to go through. The question is if the time shift between the rays of a case and the other remains constant. The answer is no. We take now $r_n(m)$ for the longest considered distance, {$d(n_f)=101\text{ m}$}. The LOS ray in the moving Tx case arrives at $\tau^{\text{Tx}}_0|_{n=n_f}=d(n_f)/c=67.33$ ms, while in the moving Rx case it arrives at $\tau^{\text{Rx}}_0|_{n=n_f}=d(n)/(c-v)\approx 69.71$ ms, so the time shift has been increased to $2.38$ ms. This time shift can be observed in all the pairs of rays from a case and the other for d(n) = 126 m. This increase of the time shift from the CIR of a case to the other, though small, shows what happens as the time goes by. This lets us formulate that the time shift from a case to the other:

\begin{equation}
\Delta m= \frac{d(n)v}{c(c-v)}, v\leq c,
\end{equation}
the further away the Rx goes the greater is the time shift with respect to the case where the Tx is the one moving away.

Finally, we would like to add that, according to (\ref{eq_basica}), the plot in Fig.  \ref{fig_particular1_tipoII} for $r^{\text{Tx}}_n(m)$ is also the plot for $p^{\text{Rx}}_n(m)$.

\subsection{Static case versus Tx and Rx moving keeping a constant distance}

Now we look at the second particular scenario introduced in subsection \ref{sss_SvsD}: The different LTI CIRs we obtain in the static case where both Tx and Rx stand still and when they both move at the same velocity (magnitude and direction), i.e. keeping the distance between them constant along time with the Tx chasing the Rx. We once again choose the distance $d=100$ m and, for the mobile case, $v_{\text{Rx}}=v_{\text{Tx}}=v=51.2$ m/s. We remark that, as detailed in \ref{sss_SvsD}, the mobile case has also a LTI CIR for the constant $v$ case, though different to the one of the static case and $v$-dependent.

\begin{figure}[!tb]
\centering
\includegraphics[width=0.75\columnwidth]{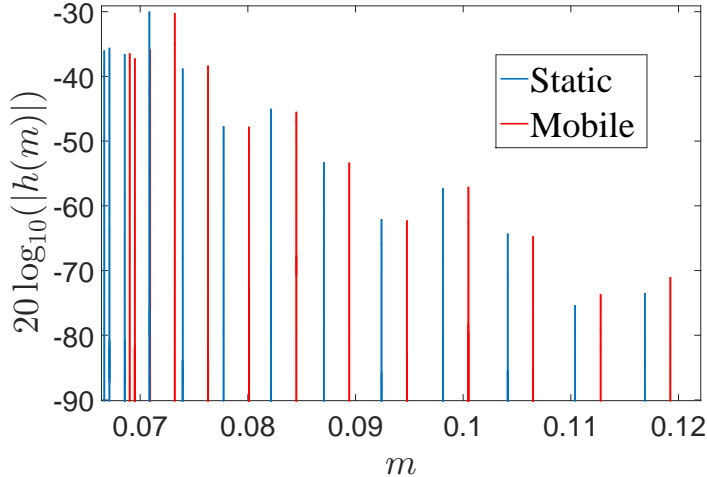}
\caption{LTI CIR for scenarios detailed in \ref{sss_SvsD}. In these examples $d=100$ m, $w=18$ m, $w_{\text{TRx}}=12$, $c_b=1300$ m/s, $\rho_b=1800$ g/m$^3$, and $v=51.2$ m/s in the dynamic case.\label{fig_particular2}}
\end{figure}

Fig. \ref{fig_particular2} show the LTI CIRs for both scenarios. Once again, we see one CIR as a shifted case of the other and the explanation can be once again found on the Rx \emph{running away} from the chasing impulse, although no Doppler effect is undergone since the channels are LTI. In fact, if we attend to the definitions in (\ref{eq_basica}), (\ref{eq_r_Rx}), (\ref{eq_static}) and (\ref{eq_dynam_LTI}), we can realize that the plots of the static  and mobile case in Fig. \ref{fig_particular2} are the plane $n=0$ of the plot of $r^{\text{Tx}}_n(m)$ and $r^{\text{Rx}}_n(m)$ respectively in Fig. \ref{fig_particular1_tipoII}.

\section{Conclusion}\label{conclusiones}

We addressed the M2M SWA channel as it was not ever before done and obtained a framework to simulate it from the static SWA channel models already available. This framework connected the M2M LTV CIR with the static LTI CIR. This work led us to find that due to the use of mechanical waves the relativity in the motion between transceivers is no longer applicable like in usual electromagnetic-wave-based communications. We illustrated this find with some numerical examples. Further development of this work should compare our model with M2M SWA sounding in scenarios where a static channel model has proved its validity.

\bibliographystyle{ieeetr}
\bibliography{bibfile}

\end{document}